\documentclass[]{aa} 

\usepackage{graphicx}
\usepackage{times}

\usepackage{txfonts}
\newcommand{\bfr}{\rm}
\begin{document}

\title{Isotopic abundances of carbon and nitrogen in Jupiter-family and
Oort Cloud comets\thanks{Based on observations collected at the
European Southern Observatory, Paranal, Chile (ESO Programme
073.C-0525).}}

\author{%
D. Hutsem\'ekers \inst{1}\fnmsep\thanks{Research Associate FNRS}\and
J. Manfroid      \inst{1}\fnmsep\thanks{Research Director FNRS} \and
E. Jehin         \inst{2}\and
C. Arpigny     \inst{1}\and
A. Cochran     \inst{3}\and
R. Schulz      \inst{4}\and
J.A. St\"uwe   \inst{5}\and
J.-M.~Zucconi  \inst{6}}

\institute{%
Institut d'Astrophysique et de G\'eophysique,
Universit\'e de Li\`ege, All\'ee du 6 ao\^ut 17, B-4000 Li\`ege
\and
European Southern Observatory, Casilla 19001, Santiago, Chile
\and
Department of Astronomy and McDonald Observatory, University 
of Texas at Austin, C-1400, Austin, USA
\and
ESA/RSSD, ESTEC, P.O. Box 299, NL-2200 AG Noordwijk, 
The Netherlands
\and
Leiden Observatory, NL-2300 RA Leiden, The Netherlands
\and
Observatoire de Besan\c{c}on, F25010 Besan\c{c}on Cedex, France
}

\date{}

\abstract{ The $^{12}$C$^{14}$N/$^{12}$C$^{15}$N\ and
$^{12}$C$^{14}$N/$^{13}$C$^{14}$N\ isotopic ratios are determined for
the first time in a Jupiter-family comet, 88P/1981 Q1 Howell, and in
the chemically peculiar Oort Cloud comet C/1999 S4 (LINEAR).  By
comparing these measurements to previous ones derived for six other
Oort Cloud comets (including one of Halley-type), we find that both
the carbon and nitrogen isotopic ratios are constant within the
uncertainties. The mean values are $^{12}$C/$^{13}$C $\simeq$ 90 and
$^{14}$N/$^{15}$N $\simeq$145 for the eight comets. These results
strengthen the view that CN radicals originate from refractory
organics formed in the protosolar molecular cloud and subsequently
incorporated in comets.
\keywords{comets: abundances -
comets: individual:  88P/Howell - 
comets: individual:  C/1999 S4 (LINEAR)
}}

   \maketitle
%

\section{Introduction}

Determination of the abundance ratios of the stable isotopes of the
light elements in different objects of the Solar System provides
important clues in the study of its origin and early history. Comets
carry the most valuable information regarding the material in the
primitive solar nebula.

The discovery of a number of emission features belonging to the
$^{12}$C$^{15}$N B-X (0,0) band ($\lambda \sim$ 3880 \AA) allowed us
to make the first optical measurement of the nitrogen isotopic ratio
$^{14}$N/$^{15}$N\ in a comet (Arpigny et al.~\cite{Arpigny}).  This
ratio ($\sim$140) was found to be lower by a factor of about two than
the terrestrial value (272) and less than half those obtained in
Hale-Bopp from millimiter measurements of HCN, a possible main parent
of CN (Jewitt et al. \cite{Jewitt}, Ziurys et al. \cite{Ziurys}).
Spectra of the fainter comets 122P/de Vico (period $\sim$74 yr) and
153P/Ikeya-Zhang (period $\sim$370 yr) gave similar results (Jehin et
al.~\cite{Jehin}). We also showed that the isotopic ratios in comets
at large heliocentric distances ($r \sim$ 3 AU) are identical within
the uncertainties to the ratios derived when the comets are closer to
the Sun (Manfroid et al. \cite{Manfroid}). These measurements are
summarized in Table~\ref{tab:all}.  The discrepancy between the
nitrogen isotopic ratios derived from CN and HCN would indicate that
cometary CN radicals are produced from at least one other source
enhanced in $^{15}$N, ruling out HCN as the major parent of CN as also
suggested by other observations (e.g. Woodney et al. \cite{Woodney}).
On the other hand, the optical determinations of the
$^{12}$C/$^{13}$C\ ratio consistently yield values around 90, only
slightly lower than the HCN millimiter measurements
(Table~\ref{tab:all}), and in agreement with the solar value (89).

As seen in Table~\ref{tab:all}, the optically determined nitrogen
isotopic ratios are remarkably similar. However, all the comets
studied so far were long-period or Halley-type comets coming from the
Oort Cloud (10$^3$--10$^5$ AU from the Sun).  Jupiter-family
short-period comets constitute a different group thought to originate
from the Edgeworth-Kuiper belt (30--10$^3$ AU from the Sun).
According to current paradigm (e.g. Weissman \cite{Weissman}),
Jupiter-family comets are believed to have formed in the
Edgeworth-Kuiper belt, although it has been argued recently that they
could have formed much closer in (Gomes \cite{Gomes}, Levison \&
Morbidelli \cite{Levison}). On the other hand, Oort Cloud comets are
in fact born in the region of the solar nebula where the giant planets
appeared (5--30 AU from the Sun).  In any case, since these two
categories of comets may have different birthplaces, it is important
to know whether their isotopic ratios differ, or not. In the present
paper, we discuss the first determination of the carbon and nitrogen
isotopic ratios in a bona-fide Jupiter-family comet: 88P/1981 Q1
Howell. It has an orbital period $P$ = 5.5 yr and a Tisserand
invariant $T_{\rm J} >$ 2 (Fernandez et al. \cite{Fernandez}).

\begin{table*}[t]
\caption{Carbon and nitrogen isotopic ratios in comets}
\label{tab:all}
\begin{tabular}{llclrrl}\hline\hline \\[-0.10in]
Comet & Type & $r$ (AU) & Method (carrier) &  $^{12}$C/$^{13}$C & $^{14}$N/$^{15}$N & References \\
       \hline \\[-0.10in]
C/1995 O1 (Hale-Bopp )   & OC    & 0.92 & Millimeter (HCN) & 109 $\pm$ 22 & 330 $\pm$ 98 & Ziurys et al.   \cite{Ziurys} \\
                         &       & 1.20 & Millimeter (HCN) & 111 $\pm$ 12 & 323 $\pm$ 46 & Jewitt et al.   \cite{Jewitt} \\[0.1cm]
C/1995 O1 (Hale-Bopp )   & OC    & 0.92 & Optical (CN)     &  90 $\pm$ 30 & 160 $\pm$ 40 & Manfroid et al. \cite{Manfroid} \\
                         &       & 0.93 & Optical (CN)     &  95 $\pm$ 40 & 140 $\pm$ 45 & Manfroid et al. \cite{Manfroid} \\
                         &       & 2.73 & Optical (CN)     &  80 $\pm$ 20 & 140 $\pm$ 30 & Manfroid et al. \cite{Manfroid} \\
C/2000 WM1 (LINEAR)      & OC    & 1.21 & Optical (CN)     & 115 $\pm$ 20 & 140 $\pm$ 30 & Arpigny et al. \cite{Arpigny} \\
C/2001 Q4 (NEAT)         & OC    & 0.98 & Optical (CN)     &  90 $\pm$ 15 & 135 $\pm$ 20 & Manfroid et al. \cite{Manfroid} \\
                         &       & 3.70 & Optical (CN)     &  70 $\pm$ 30 & 130 $\pm$ 40 & Manfroid et al. \cite{Manfroid} \\
C/2003 K4 (LINEAR)       & OC    & 1.20 & Optical (CN)     &  90 $\pm$ 15 & 135 $\pm$ 20 & Manfroid et al. \cite{Manfroid} \\
                         &       & 2.61 & Optical (CN)     &  85 $\pm$ 20 & 150 $\pm$ 35 & Manfroid et al. \cite{Manfroid} \\
122P/1995 S1 (de Vico)     & HT    & 0.66 & Optical (CN)     &  90 $\pm$ 10 & 140 $\pm$ 20 & Jehin et al. \cite{Jehin} \\
153P/2002 C1 (Ikeya-Zhang) & OC    & 0.92 & Optical (CN)     &  90 $\pm$ 25 & 170 $\pm$ 50 & Jehin et al. \cite{Jehin} \\[0.1cm]
C/1999 S4 (LINEAR)         & OC    & 0.88 & Optical (CN)     & 100 $\pm$ 30 & 150 $\pm$ 40 & This work \\
88P/1981 Q1 (Howell)       & JF    & 1.41 & Optical (CN)     &  90 $\pm$ 10 & 140 $\pm$ 15 & This work \\
\hline\\[-0.2cm]
\end{tabular}\\
{\footnotesize Comet types: OC: Oort Cloud; HT: Halley-type; JF: Jupiter-family.}
\end{table*}

Assumed to be formed at distances ranging from 5 to 30 AU from the
Sun, Oort Cloud comets may experience a variety of nebular
temperatures and densities.  Recently, Mumma et al. (\cite{Mumma})
found that the chemical composition of comet C/1999 S4 (LINEAR)
greatly differs from other Oort Cloud comets (namely Hale-Bopp),
suggesting some processing in the hotter Jupiter-Saturn region (and
then named ``Jovian-class'' Oort Cloud comet, not to be confused with
Jupiter-family comets). The chemical peculiarity of C/1999 S4 (LINEAR)
was further demonstrated by Biver et al. (\cite{Biver}) and Mumma et
al. (\cite{Mumma2}).  Comet C/1999 S4 (LINEAR) is also analysed in the
current paper, making with 88P/Howell a sample of two comets thought
to be formed in very different environments.

\section{Observations and data analysis}            

Observations of comet 88P/Howell were carried out with the
Ultraviolet-Visual Echelle Spectrograph (UVES) mounted on the 8.2m UT2
telescope of the European Southern Observatory Very Large Telescope
(ESO VLT). Eleven exposures were secured in service mode during the
period April 18, 2004 to May 24, 2004. The total exposure time amounts
to 11.1 h. The 0.44 $\times$ 8.0 arcsec slit was centered on the
nucleus and oriented along the tail, providing a resolving power $R
\simeq 80000$. The comet heliocentric distance ranges from $r$ = 1.37
AU to $r$= 1.44 AU, and its radial velocity from $\dot{r}$ = 0.89 km
s$^{-1}$ to $\dot{r}$ = 5.90 km~s$^{-1}$.

Observations of comet C/1999 S4 (LINEAR) were performed with the
2Dcoud\'e echelle spectrograph at the 2.7m Harlan J. Smith telescope
of the McDonald Observatory. The resolving power was $R \simeq
60000$. The total exposure time of 7.3 h was divided in 16 exposures
of 1200s or 1800s each, distributed in the period June 25, 2000 to
July 17, 2000, i.e. just before the comet's disruption. Heliocentric
distances and radial velocities range from $r$ = 0.97 AU to $r$ = 0.78
AU, and $\dot{r}$ = $-$19.6 km s$^{-1}$ to $\dot{r}$ = $-$7.4 km
s$^{-1}$.

\begin{figure*}[t]
\centering\includegraphics*[width=17cm]{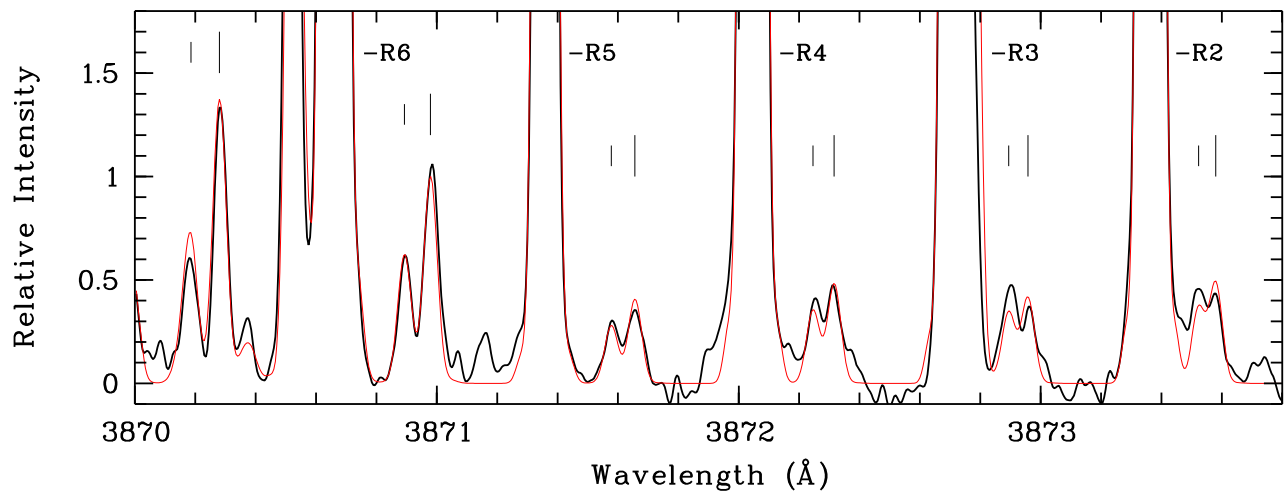}
\caption{A section of the spectrum of the CN (0,0) band in comet
88P/Howell.  {\it Thick line:} Observed spectrum; {\it thin (red)
line:} synthetic spectrum of $^{12}$C$^{14}$N, $^{12}$C$^{15}$N and
$^{13}$C$^{14}$N with the adopted isotopic abundances. The lines of
$^{12}$C$^{15}$N are identified by the short ticks and those of
$^{13}$C$^{14}$N by the tall ticks. The quantum numbers of the $R$
lines of $^{12}$C$^{14}$N are also indicated.}
\label{fig:fit}
\end{figure*}

Data reduction and analysis were done as in previous papers.
Basically, we compute synthetic fluorescence spectra of the
$^{12}$C$^{14}$N, $^{13}$C$^{14}$N and $^{12}$C$^{15}$N\
$B\,^{2}\Sigma^{+}$$-$$X\,^{2}\Sigma^{+}$ (0,0) ultraviolet bands for
each observing circumstance.  Isotope ratios are then estimated by
fitting the observed CN spectra with a linear combination of the
synthetic spectra of the three species.  For more details we refer to
Arpigny et al. (\cite{Arpigny}), Jehin et al. (\cite{Jehin}) and
Manfroid et al. (\cite{Manfroid}).  An example is shown in
Fig.~\ref{fig:fit}. The derived carbon and nitrogen isotopic ratios
are $^{12}$C$^{14}$N/$^{13}$C$^{14}$N = 90$\pm$ 10 and
$^{12}$C$^{14}$N/$^{12}$C$^{15}$N = 140$\pm$ 15 for comet 88P/Howell,
and $^{12}$C$^{14}$N/$^{13}$C$^{14}$N = 100$\pm$ 30 and
$^{12}$C$^{14}$N/$^{12}$C$^{15}$N = 150$\pm$ 40 for comet C/1999 S4
(LINEAR).  As seen from Table~\ref{tab:all}, these ratios are
identical within the uncertainties to the values we measured for
other comets.

\section{Discussion} 

The carbon and nitrogen isotopic ratios derived from CN are similar
for all the comets of our sample independently of their types and
birthplaces. The mean values are $^{12}$C/$^{13}$C = 90$\pm$4 and
$^{14}$N/$^{15}$N = 145$\pm$4 for the eight comets. This is in
remarkable contrast with the strong variability observed among other
Solar System bodies (Zinner et al. \cite{Zinner}, Owen et
al. \cite{Owen}).  {\bfr In any case, the value of the nitrogen isotope
ratio lies outside the range of measurements in the solar wind (180 --
500, Kallenbach et al. \cite{Kallenbach}) and it is inconsistent with
the lower limit ($>$ 360) derived from the study of lunar soils
(Hashizume et al. \cite{Hashizume}). It is also quite distinct from
the $^{14}$N/ $^{15}$N ratio ($\sim$450) in the atmosphere of
Jupiter, which was proposed as a protosolar value (Owen et
al. \cite{Owen}, Fouchet et al. \cite{Fouchet}).  We believe that the
ratio we measure in cometary CN is not necessarily incompatible with
the latter view, but rather suggests the possible (or likely?)
coexistence of more than one nitrogen isotope ratio characterizing
different components in the protosolar nebula.}

Since it is unlikely that all comets originate from an isolated region
of the protosolar cloud, the constancy of the $^{14}$N/$^{15}$N ratio
suggests that the isotope carrier at the origin of the CN radicals is
homegeneously distributed within the protosolar cloud and comes
unchanged from it, even in comet C/1999~S4 (LINEAR) which may have
experienced high temperatures in the Jupiter-forming region (Mumma et
al. \cite{Mumma}). This would require that CN originates from
refractory compounds aggregated in the protosolar cloud.  In fact, the
discrepancy between the nitrogen isotope ratios derived from
millimeter HCN and optical CN and the independence of the
$^{14}$N/$^{15}$N ratio on heliocentric distance has led us (Arpigny
et al. \cite{Arpigny}, Manfroid et al. \cite{Manfroid}) to suggest
that the dominant source of CN --the carrier of $^{15}$N-- could be
refractory organics. {\bfr Good candidates are HCN polymers (Rettig et
al. \cite{Rettig}, Kissel et al. \cite{Kissel}, Fray \cite {Fray}), or
grains containing organic macromolecules and reminiscent of
``cluster'' interplanetary dust particles (IDPs) known to have low
$^{14}$N/$^{15}$N ratios (Messenger \cite{Messenger}, Al\'eon et
al. \cite{Aleon}, Keller et al. \cite{Keller}).  The hypothesis
involving HCN polymers implicitly assumes that their degradation may
directly and predominantly produce CN radicals, which is not
inconsistent with current experimental limits (Fray \cite{Fray}, and
references therein). Besides, according to Rettig et
al. (\cite{Rettig}), the physico-chemical properties of HCN polymers
indicate that these compounds can likely release CN (and NH$_2$)
radicals by dissociation. The $^{14}$N/$^{15}$N ratio measured on HCN
in comet Hale-Bopp ($\sim$330) might be the result of a mixture
between unprocessed HCN with protosolar value ($\sim$450, in the sense
of Owen et al. \cite{Owen}) and $^{15}$N-enriched HCN released from
HCN polymers (characterized, as the CN we observe, by
$^{14}$N/$^{15}$N $\sim$145).}

The difference between the nitrogen isotopic ratios measured for CN
and HCN may indicate isotope fractionation in the protosolar cloud.  A
few mechanisms based on ion-molecule and gas-grain reactions in a cold
interstellar medium have been proposed to explain $^{15}$N enhancement
with respect to the isotope abundance in N$_2$ which is usually
thought to be the major nebular reservoir of nitrogen (Terzieva \&
Herbst \cite{Terzieva}, Rodgers \& Charnley \cite{Rodgers}). But it is
not clear whether such mechanisms can quantitatively reproduce the
nitrogen isotopic ratio derived from CN, its uniform distribution
throughout the protosolar cloud {\bfr whatever the fluctuations in
temperature and density}, and how it can be incorporated into
CN-bearing grains.  Moreover, the proposed mechanisms seem to be
efficient at low temperature ($\sim$10~K, Terzieva \& Herbst
\cite{Terzieva}, Charnley \& Rodgers \cite{Charnley}). Kawakita et
al. (\cite{Kawakita}), by considering the very similar spin
temperatures measured for various molecules in comets of both Oort
Cloud and Jupiter-family types, have recently proposed that the Solar
System was born in a warm dense molecular cloud near 30~K rather than
a cold dark cloud at 10~K. A higher protosolar temperature was also
suggested by Meier \& Owen (\cite{Meier}) on the basis of cometary
deuterium-to-hydrogen data.  More work is therefore needed to assess
the reality of fractionation --still very attractive-- namely in
warmer interstellar clouds.  Mass-independent nitrogen fractionation
based on selective photo-dissociation may also be worth investigating,
as done for example by Yurimoto \& Kuramoto (\cite{Yurimoto}) to
explain oxygen isotopic anomalies.

{\bfr On the other hand, the small $^{14}$N/$^{15}$N ratio we measure
in comets could originate from an external source and, for some
reason, be preferentially locked in refractory organics.
$^{15}$N~enhancement may be attributed to a contamination by
nucleosynthesis products ejected by nearby massive stars, as observed
in the Large Magellanic Cloud and starburst galaxies (Henkel \&
Mauersberger \cite{Henkel}, Chin et al. \cite{Chin}, Wang et
al. \cite{Wang}). This would be consistent with a $^{14}$N/$^{15}$N
ratio smaller than the interstellar medium (ISM) value at the solar
circle: $^{14}$N/$^{15}$N $\simeq$ 450$\pm$22 (Wang et
al.~\cite{Wang}).  A contamination by massive stars could also
slightly increase the $^{12}$C/$^{13}$C ratio with respect to
$^{12}$C/$^{13}$C $\simeq$ 77$\pm$7 measured in the local ISM (Henkel
\& Mauersberger \cite{Henkel}), as observed.  Isotopic contamination
by massive stars is independently suggested by the study of extinct
radionuclides in meteorites (e.g. Cameron et al. \cite{Cameron}).
Interestingly enough, the higher temperature of the protosolar cloud
proposed by Kawakita et al. (\cite{Kawakita}) would necessitate the
vicinity of sites of formation of massive stars.  Massive stars also
provide large amount of ultraviolet radiation which may initiate HCN
polymerization as suggested by Rettig et al. (\cite{Rettig}). If this
polymerization (or another, unknown, mechanism) locked up the
$^{15}$N-enriched nitrogen in a solid phase before it spreads
throughout the protosolar cloud, this could explain the nitrogen
isotopic differences observed between different molecules in comets
and among the various objects and reservoirs in the Solar System (Owen
et al. \cite{Owen}).  A similar effect could be expected for the
carbon isotopes although much smaller and within the errors bars.}

{\bfr Apart from the need to identify contamination or fractionation
processes appropriate to either type of CN progenitors, polymers or
organic macromolecules, it will also be very important to measure the
$^{14}$N/$^{15}$N ratio in different N-bearing molecules to establish
a complete inventory of nitrogen isotopes in comets: HCN (for which a
single measurement has been possible so far), NH$_3$ and/or NH$_2$
(which is a direct product of ammonia), N$_2$, C$_2$N$_2$ if present.
Some variation may be expected depending on the mechanisms of $^{15}$N
enrichment and the sizes of the various nitrogen reservoirs.  A
comparison with accurate measurements of the solar-wind isotopes from
samples collected by the Genesis space mission should help to
distinguish between the possible scenarios.}

\end{document}